# On the masses of neutron stars

**M.H. van Kerkwijk**[1,2], **J. van Paradijs**[1,3], **and E.J. Zuiderwijk**[4,5]

[1] Astronomical Institute "Anton Pannekoek", University of Amsterdam, and Center for High-Energy Astrophysics (CHEAF), Kruislaan 403, 1098 SL Amsterdam, The Netherlands
[2] Department of Astronomy, California Institute of Technology, m.s. 105-24, Pasadena, CA 91125, USA
[3] Physics Department, University of Alabama in Huntsville, Huntsville, AL 35899, USA
[4] Kapteyn Laboratorium, Zernike Gebouw, Postbus 800, 9700 AV Groningen, The Netherlands
[5] Royal Greenwich Observatory, Madingley Road, Cambridge CB3 0EZ, United Kingdom

**Abstract.** We analyze the currently available observations of X-ray binaries in a consistent way, to re-determine the masses of the neutron stars in these systems. In particular, our attention is focussed on a realistic and consistent assessment of observational uncertainties and sources of systematic error. Confidence limits for these new mass estimates are generally less optimistic than previously assumed. The available observations, including data on six radio pulsars, do not firmly constrain the equation of state of neutron star matter. In particular, a firm upper mass limit cannot yet be established. An improvement of the accuracy of optical data holds the key to further progress.

## 1. Introduction

X-ray and radio pulse-timing observations of neutron stars have provided most of our knowledge on the structure of these objects, and on the properties of the dense matter of which they are composed (for general references, see the NATO ASI Series C, Volumes 262, 344 and 377). A basic property of neutron stars accessible to observational study, is their mass or, more generally, their mass-radius relation. To a large extent, this relation, and in particular the maximum possible mass ($M_{\rm max}$) for a (stable) neutron star, is determined by the equation of state of condensed matter. For "soft" equations of state $M_{\rm max} \simeq 1.4\,M_\odot$, and $R \lesssim 10\,{\rm km}$, while for very "stiff" equations of state the values are up to $\sim 1.8\,M_\odot$, and $\sim 15\,{\rm km}$, respectively (e.g., Shapiro & Teukolsky 1983; Arnett & Bowers 1977; Datta 1988; Stock 1989; Brown & Bethe 1994).

Attempts have been made to find constraints on the mass-radius relation from, for example, X-ray burst observations (see Lewin et al. 1993), from the limiting spin period of neutron stars (Friedman et al. 1986), from their cooling history (e.g., Ögelman 1995; Umeda et al. 1993), from accretion-induced spin-up/spin-down behaviour combined with magnetic-field strength estimates from cyclotron lines (Wasserman & Shapiro 1983), from glitches in radio pulsars (Alpar 1995; cf. Lorenz et al. 1993), and from the neutrino intensity curve during a supernova (Loredo & Lamb 1989). So far, however, none of these methods have led to useful constraints on the mass-radius relation.

Direct mass estimates have been obtained for 17 neutron stars in binary systems, with varying degree of accuracy. Four of these estimates are either only a limit to the mass or have otherwise large observational errors. The remaining 13 objects comprise six X-ray pulsars in binaries with an observable optical companion, and four millisecond radio pulsars in close binary systems, of which three have a neutron-star companion.

For the radio-pulsar binaries, the masses (see Table 1) of both components can be determined from an analysis of the pulse arrival times, if one assumes that general relativity correctly describes the gravitational interaction in these systems (e.g., Taylor & Weisberg 1989). The orbital period, eccentricity and rate of periastron advance provide one relation between the two masses (their sum), while the orbital variations of the gravitational redshift and the transverse Doppler shift provide another one. Alternatively, if the Shapiro delay can be measured, one obtains constraints on both the inclination and the mass of the companion, and thus, using the mass function, also on the mass of the pulsar (Ryba & Taylor 1991).

For the six X-ray binaries, the system parameters can, in principle, also be determined completely, because both the X-ray and optical orbits can be measured, and information about the orbital inclination is contained in the X-ray eclipse duration (combined with an estimate of the size of the optical component). Rappaport & Joss (1983) reviewed these neutron-star masses, and a (partial) update was given by Nagase (1989). Then, as well as now, the uncertainty of the mass determinations

---

*Send offprint requests to*: M.H. van Kerkwijk (Caltech)

**Table 1.** Parameters for the radio pulsars[a]

| PSR | $P_{\rm orb}$ (day) | $M_{\rm total}$ ($M_\odot$) | $M_{\rm pulsar}$ ($M_\odot$) | $M_{\rm comp.}$ ($M_\odot$) | Ref. |
|---|---|---|---|---|---|
| 1913+16 | 0.323 | 2.82837(4) | 1.442(3) | 1.386(3) | 1 |
| 1534+12 | 0.421 | 2.679(3) | 1.32(3) | 1.36(3) | 2 |
| 2127+11C | 0.335 | 2.712(5) | 1.34(23) | 1.37(23) | 3 |
| 1855+09[b] | 12.33 | | $1.27^{+0.23}_{-0.15}$ | $0.233^{+0.026}_{-0.017}$ | 4 |

[a] Numbers in parentheses and errors indicate $1\sigma$ confidence limits. References: (1) Taylor & Weisberg 1989; (2) Wolsczcan 1991; (3) Anderson 1993; (4) Ryba & Taylor 1991
[b] The total mass is not independently determined for this system; the companion is not a neutron star

was dominated by the relative inaccuracy of the optical data. However, optical observing techniques have improved considerably over the past decade, and much new material has become available. Therefore, a fresh assessment of those six currently available mass estimates seems to be called for[1]. In this paper, we present such an assessment.

## 2. Six X-ray binaries revisited

We present this overview as a follow-up on optical work on Vela X-1 (Van Kerkwijk et al. 1995, hereafter Paper I). For this system, we found that an accurate determination of the optical orbit was hampered both by the presence of pulsation-like perturbations and by the possible presence of systematic effects occurring at a specific orbital phase. The two combined limit the accuracy of the measured radial-velocity amplitude to a 95% confidence range of $18.0$–$28.2\,{\rm km\,s^{-1}}$, in spite of the fact that individual velocities can now be measured to better than $1\,{\rm km\,s^{-1}}$.

The optical radial-velocity amplitude, $K_{\rm opt}$, is one of the three least known parameters, the other two being the duration of the X-ray eclipse, $\theta_{\rm ecl}$, and the projected rotational velocity, $v\sin i$, of the optical star. We have, therefore, focussed on these three quantities.

The uncertainty in $K_{\rm opt}$ depends, apart from the observational errors, on the possible presence of orbital and non-orbital perturbations, such as in Vela X-1, and on systematic effects arising from the tidal distortion of the companion, heating by X rays, and the presence of an accretion disk. The perturbations in Vela X-1 might be related to the eccentricity of the orbit (Paper I). However, the other effects could also be present in circular systems. In Vela X-1 their contribution is $\lesssim 5\%$ (Paper I), an upper limit which we will consider as typical for a system without a prominent disk and X-ray heating. In systems that do have these attributes, however, the effects may be much more important.

For the X-ray eclipse duration we use, if possible, observations made in hard X-rays, since at lower energies the eclipses may be systematically longer due to absorption in the stellar wind of the companion. The different determinations are often not consistent with each other (in Vela X-1 this may be related to variable distortions of the shape of the companion). Therefore, we have used a range within which the duration almost certainly lies ($\sim 99\%$ confidence), rather than a number with a corresponding '$1\sigma$ error'. The same approach was applied to the measured (projected) rotational velocity, for those cases where individual determinations often were inconsistent.

Below, we discuss the individual systems in detail. We give a reference to the most recent determination of the X-ray orbit right after the source name. The results are summarised in Table 2.

*Vela X-1.* (Deeter et al. 1987) This system is discussed in detail in Paper I. Briefly, for $K_{\rm opt}$ we use the 99% confidence range from that paper, for $\theta_{\rm ecl}$ a range based on work by Watson & Griffiths (1977; $33°\!\!.8 \pm 1°\!\!.3$, ARIEL V), Nagase et al. (1983; $32° \pm 1°$, Hakucho 9–22 keV) and Sato et al. (1986; $34°\!\!.4 \pm 1°\!\!.1$, Tenma 10–20 keV), and for $v\sin i$ the result of Zuiderwijk (1995; $116 \pm 6\,{\rm km\,s^{-1}}$).

*4U 1538−52.* (Makishima et al. 1987; Cominsky & Moraes 1991) The range of $27°$–$30°$ for $\theta_{\rm ecl}$ is indicated by results from Ginga ($28°\!\!.3 \pm 0°\!\!.5$; Corbet et al. 1993), Tenma ($25° \pm 5°$; Makishima et al. 1987), OSO-8 ($30°\!\!.5 \pm 3°\!\!.5$; Becker et al. 1977) and ARIEL V ($28° \pm 3°$; Davison et al. 1977).

A recent determination of the optical radial-velocity curve was made by Reynolds et al. (1992). They present two values of $K_{\rm opt}$, one as measured directly ($19.2 \pm 1.2\,{\rm km\,s^{-1}}$), and one for the velocities corrected for tidal distortion of the star ($19.8 \pm 1.1\,{\rm km\,s^{-1}}$). Such corrections are always model dependent, with a corresponding substantial uncertainty (see Paper I). The situation is further complicated by the poor phase coverage of the observations, which does not allow one to test other predicted effects such as a spurious eccentricity. In fact, the system may well be genuinely eccentric ($e = 0.08 \pm 0.05$; Makishima et al. 1987), and systematic velocity excursions like the ones observed in Vela X-1 might be present. In this respect, we note that the profiles of the line used for the velocity determination (He I $\lambda 6678$) show systematic changes similar to those seen in Vela X-1. If such is the case, the observations taken during one night cannot be treated as being independent (Paper I), and the actual error on the radial-velocity amplitude would be larger by about a factor $\sqrt{n_{\rm s}/n_{\rm n}}$, where $n_{\rm s}$ is the number of spectra and $n_{\rm n}$ the number of nights (i.e., a factor of $\gtrsim \sqrt{5}$).

An uncertainty of a different kind in the work of Reynolds et al. (1992), is that the velocities are derived from cross-correlations of the spectra with the average spectrum. If proper precautions are not taken this may result in a systematic *un-*

---

[1] We have excluded a priori the mass determinations made for 4U 1700-37 ($M_{\rm X} = 1.8 \pm 0.4\,M_\odot$, Heap & Corcoran 1992) and 4U 1626−67 ($M_{\rm X} = 1.8^{+2.9}_{-1.3}\,M_\odot$, Middleditch et al. 1981). For 4U 1700-37, the determination is based on a mass of the optical companion estimated from on its spectral type, which, in our opinion, introduces a large uncertainty that may well invalidate the quoted error. For 4U 1626−67, the orbital period found in the optical pulsations has not yet been confirmed by X-ray observations (Levine et al. 1988), and a meaningful mass determination is not yet possible.

**Table 2.** Orbital parameters of the X-ray pulsars[a]

| Name | $T_0$ (JD-2440000) | $P_{\rm orb}$ (day) | $\dot{P}_{\rm orb}/P_{\rm orb}$ (yr$^{-1}$) | $a_X \sin i$ (lt-s) | $e$ | $K_{\rm opt}$ (km s$^{-1}$) | $v \sin i$ (km s$^{-1}$) | $\theta_{\rm ecl}$ (°) |
|---|---|---|---|---|---|---|---|---|
| Vela X-1[b] | 4279.0466(37) | 8.964416(49) | $< 1.9\,10^{-5}$ | 112.98(35) | 0.0885(25) | 17.0–29.7 | 116(6) | 30–36 |
| 4U 1538−52 | 7221.974(20) | 3.72844(2) | $< 1.2\,10^{-5}\,(3\sigma)$ | 50.9(35) | 0.08(5) | 20(3) | 140–220 | 27–30 |
| SMC X-1 | 2836.68277(20) | 3.89229118(48) | $-3.36(2)\,10^{-6}$ | 53.4876(4) | $< 0.00004$ | 23(3) | 130–220 | 27–31 |
| LMC X-4 | 7742.4904(2) | 1.40839(1) | $< 3\,10^{-6}$ | 26.31(3) | $< 0.01$ | 38(5) | 120–220 | 24–29 |
| Cen X-3 | 958.8509(3) | 2.0871390(9) | $-1.78(8)\,10^{-6}$ | 39.636(3) | $< 0.0008$ | 24(6) | 250(30) | 31–37 |
| Her X-1 | 3805.019980(14) | 1.700167720(10) | $-1.32(16)\,10^{-8}$ | 13.1831(3) | $< 0.0003$ | 90(20) | | 24–25 |

[a] Numbers in parentheses indicate $1\sigma$, ranges approximate 99%, and upper limits $2\sigma$ confidence limits, except when indicated otherwise.
[b] $T_0$ refers to the time of mean longitude $90°$; periastron angle $\varpi = 150°\!\!.6 \pm 1°\!\!.8$; $|\dot{\varpi}| < 1°\!\!.9\,{\rm yr}^{-1}\,(2\sigma)$

*der*estimation of $K_{\rm opt}$, because of the presence of a component in the cross-correlation peak that is due to autocorrelated noise (see Paper I). For the rather noisy spectra involved, we estimate that this systematic effect could be as large as 10%.

In order to account for all uncertainties mentioned above, we adopt $K_{\rm opt} = 20 \pm 3\,{\rm km\,s}^{-1}$. For $v \sin i$, we use a range of 140–220 km s$^{-1}$, based on the determinations made by Crampton et al. (1978; $200 \pm 20\,{\rm km\,s}^{-1}$) and by Reynolds et al. (1992; $160 \pm 10\,{\rm km\,s}^{-1}$ plus a possible systematic error).

*SMC X-1.* (Levine et al. 1993) The values for $\theta_{\rm ecl}$, $28°\!\!.2 \pm 0°\!\!.9$ ('extreme' limits; SAS-3; Primini et al. 1976) and $29°\!\!.9 \pm 0°\!\!.2$ (COS-B; Bonnet-Bidaud & Van der Klis 1981), appear only marginally consistent. Therefore, we adopt a range of 27°–31°.

The radial-velocity curve was recently re-determined by Reynolds et al. (1993). Like for 4U 1538−52, they give both an 'uncorrected' value of $K_{\rm opt}$ ($23.0 \pm 1.9\,{\rm km\,s}^{-1}$), and a value corrected for tidal effects and X-ray heating ($27.5 \pm 1.9\,{\rm km\,s}^{-1}$). This correction is more problematic than that for 4U 1538−52, because, as the authors point out themselves, their model does not allow for the presence of an accretion disk, whereas analysis of the optical light curve shows that a disk is almost certainly present (e.g., Tjemkes et al. 1986). Its shadow may well reduce the effect of X-ray heating. Also, the size (and even the sign) of any heating effect strongly depends on the unknown presence or absence of a substantial soft X-ray flux. Therefore, we have taken a rather cavalier approach to such uncertainties and adopted a bottom-line relative error of 10% to account for possible systematic effects, i.e., we use $K_{\rm opt} = 23 \pm 3\,{\rm km\,s}^{-1}$. Notice that the orbit of SMC X-1 is circular, and that no systematic velocity excursions are observed (nor are they expected).

The range in $v \sin i$ of 130–220 km s$^{-1}$ encompasses the estimates of Hutchings et al. (1977; $\sim 200\,{\rm km\,s}^{-1}$) and Reynolds et al. (1993; $\sim 150\,{\rm km\,s}^{-1}$ with a possible systematic error).

*LMC X-4.* (Levine et al. 1991) The value of $\theta_{\rm ecl}$ is listed as $29°\!\!.0 \pm 2°\!\!.5$ by Li et al. (1978; SAS-3, 6–12 keV), $26°\!\!.2 \pm 1°\!\!.1$ by White (1978; ARIEL V), and $27°\!\!.1 \pm 1°\!\!.0$ by Pietsch et al. (1985; EXOSAT). Thus, a range of 24°–29° seems indicated.

Kelley et al. (1983a) derived $K_{\rm opt} = 37.9 \pm 2.4\,{\rm km\,s}^{-1}$ from measurements by Hutchings et al. (1978) and Petro & Hiltner (unpublished). For the mass estimates, they set the error to 5 km s$^{-1}$, in order to account for possible systematic effects due to tidal distortion and X-ray heating (both are seen in the optical light curve; Heemskerk & Van Paradijs 1989). With the estimate of 10% relative uncertainty made for SMC X-1, we find the same value.

Hutchings et al. (1978) estimated $v \sin i \simeq 170\,{\rm km\,s}^{-1}$. Based on the variations between different estimates for the other sources, we conservatively use a range of 120–220 km s$^{-1}$. Notice that if this value were correct, the star would be rotating subsynchronously (Table 3). It seems worthwhile to make a more accurate determination.

*Cen X-3.* (Kelley et al. 1983b) Clark et al. (1988; SAS-3 observations) found that for different eclipses, $\theta_{\rm ecl}$ ranges from 33° to 37° at 10–20 keV, while at 3–6 keV a range of 35°–40° is indicated. (For comparison, Pounds et al. (1975; ARIEL V) give $39° \pm 2°$, and Schreier et al. (1972; UHURU) find $42° \pm 1°$). From fits to the eclipses in different energy bands with a simple wind model, they found that the "real" value of $\theta_{\rm ecl}$ should be in the range 31°–36°. Conservatively, we have adopted a range 31°–37°.

Hutchings et al. (1979) derived $K_{\rm opt} = 24 \pm 6\,{\rm km\,s}^{-1}$. Given the large error, it is immaterial whether we account for possible systematic effects. We note that the optical light curve indicates the likely presence of an accretion disk, but does not show strong evidence for X-ray heating (Tjemkes et al. 1986). For $v \sin i$, we used the result of Hutchings et al. (1979): $250 \pm 30\,{\rm km\,s}^{-1}$.

*Her X-1.* (Deeter et al. 1981, 1991) Deeter et al. (1981; OSO-8) determined a very precise value of $\theta_{\rm ecl}$, $24°\!\!.56 \pm 0°\!\!.03$. However, this value is based on an ingress and an egress of different orbital cycles. Also, their Copernicus results indicate a somewhat larger range in $\theta_{\rm ecl}$. Therefore, we used a range 24°–25°. We were unable to find a limit on $v \sin i$. Following Rappaport & Joss (1983), we used a range of 0–1.5 for the corotation factor.

Two mass determinations have been published for this source. One uses $K_{\rm opt}$, like for the other sources. For this, Hutchings et al. (1985) find $83 \pm 3\,{\rm km\,s}^{-1}$. However, this error reflects solely the observational one, and not the one associated with the (model-dependent) correction for the tidal deformation and especially the intense X-ray heating (known to vary with the 35-day precession period). The latter is likely large. For

4    M.H. van Kerkwijk et al.: On the masses of neutron stars

**Table 3.** Inferred parameters for the X-ray pulsars[a]

| Name | $f_{\rm co}$ | $i$ (°) | $a$ ($R_\odot$) | $R_{\rm opt}$ ($R_\odot$) | $M_{\rm opt}$ ($M_\odot$) | $M_{\rm X}$ ($M_\odot$) | $f_{\rm bad}$[b] (%) |
|---|---|---|---|---|---|---|---|
| Vela X-1 | $0.69^{+0.09}_{-0.08}$ | $>74$ | $53.4^{+1.6}_{-1.4}$ | $30.0^{+1.8}_{-1.9}$ | $23.5^{+2.2}_{-1.5}$ | $1.88^{+0.69}_{-0.47}$ | 44 |
| 4U 1538−52[c] | $0.94^{+0.32}_{-0.25}$ | $68^{+9}_{-7}$ | $26.2^{+2.5}_{-2.3}$ | $15.3^{+2.8}_{-2.6}$ | $16.4^{+5.2}_{-4.0}$ | $1.06^{+0.41}_{-0.34}$ | 0 |
| SMC X-1 | $0.95^{+0.34}_{-0.27}$ | $70^{+11}_{-7}$ | $26.4^{+1.4}_{-1.3}$ | $15.0^{+2.3}_{-2.1}$ | $15.2^{+2.3}_{-2.1}$ | $1.17^{+0.36}_{-0.32}$ | 0 |
| LMC X-4 | $0.65^{+0.23}_{-0.19}$ | $65^{+7}_{-6}$ | $13.7^{+0.6}_{-0.6}$ | $8.0^{+1.0}_{-0.9}$ | $15.8^{+2.3}_{-2.0}$ | $1.47^{+0.44}_{-0.39}$ | 0 |
| Cen X-3 | $0.95^{+0.27}_{-0.25}$ | $>66$ | $18.7^{+1.2}_{-0.7}$ | $11.1^{+1.8}_{-1.1}$ | $18.9^{+4.0}_{-1.8}$ | $1.09^{+0.57}_{-0.52}$ | 7 |
| Her X-1 | 0.0–1.5[d] | $>72$ | $8.7^{+1.1}_{-1.1}$ | $3.93^{+0.27}_{-0.52}$ | $2.04^{+0.49}_{-0.45}$ | $1.04^{+0.75}_{-0.58}$ | 18 |
| Her X-1[e] | 0.0–1.5[d] | $>79$ | $9.3^{+0.3}_{-0.6}$ | $4.01^{+0.23}_{-0.37}$ | $2.32^{+0.16}_{-0.29}$ | $1.47^{+0.23}_{-0.37}$ | 32 |

[a] Errors and lower limits are 95% confidence
[b] Fraction of trials rejected because eclipse width could not be fit; for a discussion, see Paper I
[c] For a circular orbit. For $e = 0.08(5)$ (Table 2) and $\varpi$ from a range 0–$2\pi$, $M_X = 0.96^{+0.38}_{-0.32}$
[d] Fixed. No information on the rotational velocity is available
[e] Results for the optical pulsation Doppler-shift amplitude of $20.0 \pm 2.4\,{\rm km\,s^{-1}}$ (see Sect. 2)

instance, Koo & Kron (1977) found a value of $110 \pm 17\,{\rm km\,s^{-1}}$ after correction of their data. To accommodate this large uncertainty, we adopt $K_{\rm opt} = 90 \pm 20\,{\rm km\,s^{-1}}$.

The other mass estimate is based on an analysis of the optical pulses that are caused by reprocessing of the (pulsed) X rays in the accretion disk and on the surface of the optical star (Middleditch & Nelson 1976; Middleditch 1983). With a geometrical model for the reprocessing regions (Middleditch & Nelson 1976; Bahcall & Chester 1977), one can derive the orbital parameters and limits on the masses of the two components from the Doppler-shift amplitude of the optical pulses ($20.0 \pm 1.4\,{\rm km\,s^{-1}}$) combined with the X-ray orbit and the duration of the X-ray eclipse. We apply (again) an additional relative uncertainty of 10% to account for the uncertainty in the site where the reprocessing takes place (i.e., $20.0 \pm 2.4\,{\rm km\,s^{-1}}$). This mass determination appears to be more accurate than the one based on $K_{\rm opt}$. The largest contribution to the error results from the uncertainty in the corotation factor. If a value 1.0 is assumed, instead of a range 0–1.5, the 95% confidence error is reduced to $0.12\,M_\odot$ from the current $\sim 0.3\,M_\odot$.

## 3. Results and discussion

With the observed parameter ranges (Table 2), we determined the corotation factor, inclination, semi-major axis, radius and mass of the optical component, and mass of the neutron star, by means of Monte-Carlo simulations similar to those described by Rappaport & Joss (1983). Following these authors, we used a range 0.95–1.0 for the filling factor $\beta$ in Her X-1 (for a justification, see Bahcall & Chester 1977) and 0.9–1.0 for the other systems, for which it is known from the analysis of the optical light curves that they are close to filling their Roche lobe (e.g., Tjemkes et al. 1986). The results are listed in Table 3.

The neutron-star masses derived for both X-ray and radio pulsars are shown in Fig. 1. Clearly, the data are consistent with a narrow range of neutron star-masses, with the upper limit set at $1.44\,M_\odot$ by PSR 1913+16 and the lower at $1.36\,M_\odot$ by PSR 2127+11C or its companion (it is not clear which one; the limit results from the fact that the average mass of the system is well constrained). However, it is also obvious that a much wider range of neutron star masses is very well possible (see, e.g., Finn 1994 for a constraint derived from the radio determinations). An upper limit of $1.44\,M_\odot$ does not provide strong constraints on the equation of state. Only 'the ideal neutron-star gas', equation of state 'H' as listed by Arnett & Bowers (1977), can be ruled out (as was already known), while the very soft equations of state 'G' and 'B' are likely to be unrealistic.

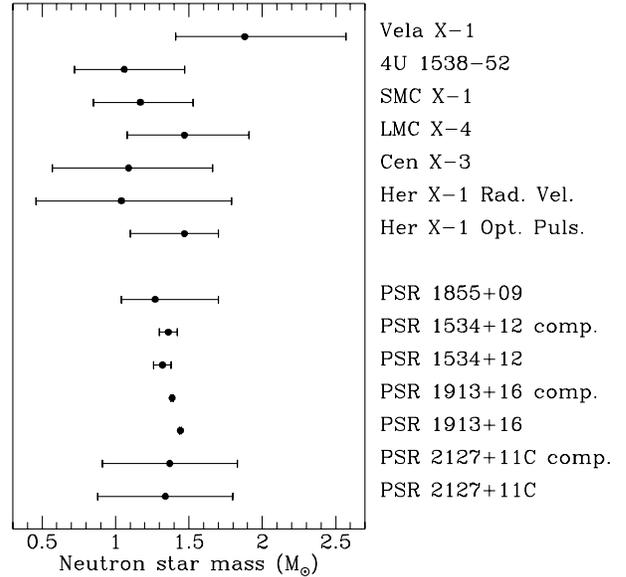

**Fig. 1.** The currently known masses of neutron stars. Shown are both the masses for the X-ray pulsars (Table 3) and for the radio pulsars (and their companions; Table 1). The error bars indicate 95% confidence limits. For the radio pulsars, these have been obtained by multiplying the $1\sigma$ errors listed in Table 1 by two (except for PSR 1855+09, for which we read $M_{\rm pulsar} = 1.27^{+0.43}_{-0.23}$ from Fig. 8 of Ryba & Taylor 1991)

Clearly, more accurate determinations are necessary. For the radio pulsars, this refers to the quantity of determinations, while for the X-ray pulsars it is the quality that needs improvement. The lack of the latter still mainly results from the uncertainties in the three parameters mentioned earlier, the radial-velocity amplitude $K_{\rm opt}$ (for all systems), the projected rotational velocity $v \sin i$ (in particular for Her X-1) and the eclipse duration $\theta_{\rm ecl}$ (for Vela X-1 and Cen X-3). An improvement of the accuracy of these quantities by a factor of two would already greatly improve the situation. For $v \sin i$, this would require only a minor effort, while improved determinations for $\theta_{\rm ecl}$ are likely to become available from, e.g., the analysis of the BATSE database (Chakrabarty et al., in preparation). However, for $K_{\rm opt}$ the situation is less hopeful. Our study of Vela X-1 (Paper I) has shown that at least for that system it will require a great effort to obtain a significant improvement, both observationally and theoretically.

*Acknowledgements.* We thank Thomas Augusteijn, Alan Levine and Saul Rappaport for help with the Monte-Carlo simulations. MHvK acknowledges support by NASA through a Hubble fellowship (HF-1053.01-93A) awarded by the Space Telescope Science Institute.